# Volumetric Reconstruction of Prostatectomy Specimens from Histology

Authors: Tom Bisson[1,2], Isil Dogan O[2], Iris Piwonski[1], Tim-Rasmus Kiehl[1,2], Georg Lukas Baumgärtner[3], Rita Carvalho[4], Peter Hufnagl[1,2], Tobias Penzkofer[3], Norman Zerbe[1,2,5], Sefer Elezkurtaj[1]

1 Charité – Universitätsmedizin Berlin, corporate member of Freie Universität Berlin and Humboldt Universität zu Berlin, Institute of Pathology

2 Charité – Universitätsmedizin Berlin, corporate member of Freie Universität Berlin, Humboldt Universität zu Berlin, Institute of Medical Informatics

3 Charité – Universitätsmedizin Berlin, corporate member of Freie Universität Berlin, Humboldt Universität zu Berlin, Department of Radiology

4 Department of Pathology, MVZ Helios Klinikum Emil von Behring, Berlin, Germany

5 Institute of Pathology, University Hospital RWTH Aachen, Aachen, Germany

# Abstract

Objectives

Surgical treatment for prostate cancer often involves organ removal, i.e., prostatectomy. Pathology reports on these specimens convey treatment-relevant information. Beyond these reports, the diagnostic process generates extensive and complex information that is difficult to represent in reports, although it is of significant interest to the other medical specialties involved. 3D tissue reconstruction would allow for better spatial visualization, as well as combinations with other imaging modalities. Existing approaches in this area have proven labor-intensive and challenging to integrate into clinical workflows. 3D-SLIVER provides a simplified solution, implemented as an open-source 3DSlicer extension. We outline three specific real-world scenarios to illustrate its potential to improve transparency in diagnostic workflows and contribute to multi-modal research endeavors.

Methods

Implementing the 3D reconstruction process involved four sub-modules of 3D-SLIVER: digitization of slicing protocol, virtual slicing of arbitrary 3D models based on that protocol, registration of slides with virtual slices using the Coherent Point Drift algorithm, and 3D reconstruction of registered information using convex hulls, Gaussian splatter and linear extrusion.

Results

Three use cases to employ 3D-SLIVER are presented: a low-effort approach to pathology workflow integration and two research-related use cases illustrating how to perform retrospective evaluations of PI-RADS predictions and statistically model 3D distributions of morphological patterns.

Conclusions

3D-SLIVER allows for improved interdisciplinary communication among specialties. It is designed for simplicity in application, allowing for flexible integration into various workflows and use cases. Here we focused on the clinical care of prostate cancer patients, but future possibilities are extensive with other neoplasms and in education and research.

# Introduction

Prostate cancer (PCa), i.e., prostatic adenocarcinoma, one of the most common cancers in men worldwide, presents a complex and multifaceted challenge for healthcare systems. Successful management necessitates a multidisciplinary approach requiring close collaboration between the various disciplines. In this context, urology, oncology, radiology, and pathology play a central role in the screening, diagnosis, and treatment of PCa.

When a patient is suspected to have PCa, a biopsy is performed by urology and sent to pathology. If the biopsy confirms the presence of PCa, this patient will, in many cases, subsequently undergo radical prostatectomy. The goal here is to resect a localized tumor entirely or to remove a significant portion of a progressed or metastasized cancer. Oncologists, specialized in cancer treatment, play a key role in the medical care of PCa patients. They are instrumental in recommending and coordinating various therapeutic approaches, such as chemotherapy, radiation therapy, hormone therapy, immunotherapy, or radical prostatectomy. Close collaboration with other medical disciplines is critical to developing and monitoring an individualized treatment strategy for individual patients. Radiologists use imaging techniques to detect PCa and determine its clinical stage. Transrectal ultrasonography (TRUS) and magnetic resonance imaging (MRI) are typically the methods of choice for this tumor type. A significant tool in this context is the Prostate Imaging Reporting and Data System (PI-RADS), [1] which is based on multiparametric MRI, a series of different sequences to estimate the likelihood of PCa in specific anatomic sectors. Pathology is central to the evaluation of prostate tissue. Pathologists examine both biopsies and radical prostatectomies to confirm the presence of neoplasia and determine the grade of PCa. This information is critical for choosing appropriate treatment or adjuvant therapy options. In addition to the Gleason grade, pathological tumor staging and resection margin status are also provided where possible.

Joint communication and collaboration between pathology and other medical disciplines is very important. In oncology, diagnostic results are needed as they indicate therapy, allow monitoring of therapy response, and contribute to predicting the long-term outcome. In radiology, diagnostic results provide valuable feedback to the predictions and may therefore improve the quality of radiological interpretation in the long term. In urology, biopsy results can be used to support surgical planning. They might also be incorporated in retrospective evaluations of surgical procedures by providing information on the extent of tumor removal and indicating cases where tissue may have been removed unnecessarily.

In pathology laboratories, tissue sample preparation for the microscopic examination involves several steps: fixation, paraffin embedding, microtomy and staining. These steps can introduce many artifacts [2]. Depending on the fixation solvent and duration used, the tissue may shrink or swell, which in turn may lead to some deformation of individual structures. Furthermore, inadequate fixation may even compromise the integrity of the tissue. Damage to the tissue can occur during the histological sectioning process itself, where, among others, fragmentation into multiple pieces or tissue cracks can occur. Finally, the tissue may even wrinkle or fold when the tissue is put onto the glass slide or during staining.

The problem of tissue deformation is by no means trivial when the objective is to combine histopathology and radiology [3]. A wide range of studies have already been carried out, and various approaches to solving this problem have been presented. These include introducing artificial reference points into the resected tissue before histologic sectioning [4], aligning the angle of sectioning to the MRI image slices [5, 6], or even performing MRI ("ex vivo") imaging of the already-resected and fixed tissue to obtain an intermediate registration [7]. An additional hurdle is that the longitudinal tissue sections of the prostate need to be further subdivided to fit onto standard glass slides, introducing additional tissue deformations compared to the MRI. Again, this problem has already been solved by photographing the longitudinal sections before further dissection, thus also allowing for intermediate registration [8]. Finally, approaches have been presented in which customized compression molds for the histological tissue preparation based on the MRI images are manufactured using a 3D printer [9, 10, 11, 12]. These allow for sectioning that has perfect alignment with the MRI, which can again significantly increase the quality of registration. However, all of these methods require significant effort to be implemented and thus are difficult to integrate into the daily routine workflow of a pathology laboratory.

In this publication, we describe a process that combines the histological evaluation with 3D prostate models. We aim to contribute to an increased exchange between all involved disciplines, as this allows sharing much more information than the final pathology report. Since the previously introduced methods are laborious and difficult to integrate into clinical routine, our solution allows for a 3D reconstruction of prostatectomy specimens with little manual effort, including localization and extension of individual tumor foci. To make this workflow accessible to the scientific community, we implemented 3D-SLIVER, an extension for the open-source software 3D Slicer [13]. It is designed to work with 3D models from arbitrary sources, such as MRI, 3D scanners, photogrammetry or even a generic model of the prostate. It thus allows for a wide range of applications.

## Materials and Methods

This 3D reconstruction of prostatectomy specimens comprises four sub-steps implemented as independent modules within 3D-SLIVER. All intermediate results can be exported or even generated in external workflows to allow for a flexible implementation to user-specific needs. The reconstruction is based on a 3D reference model into which the results of the histopathological diagnostic images are transferred through registration, followed by volume reconstruction. For this purpose, a 3D surface model of the respective prostate is divided into individual polygons that are analogous to the dissection of the prostatectomy specimen and correspond to the respective histological sections. The 3D surface model can be generated using various techniques, such as 3D scanners, photogrammetry or segmentation from MRI images. Figure 1 gives an overview of this process using the different modules and their interplay.

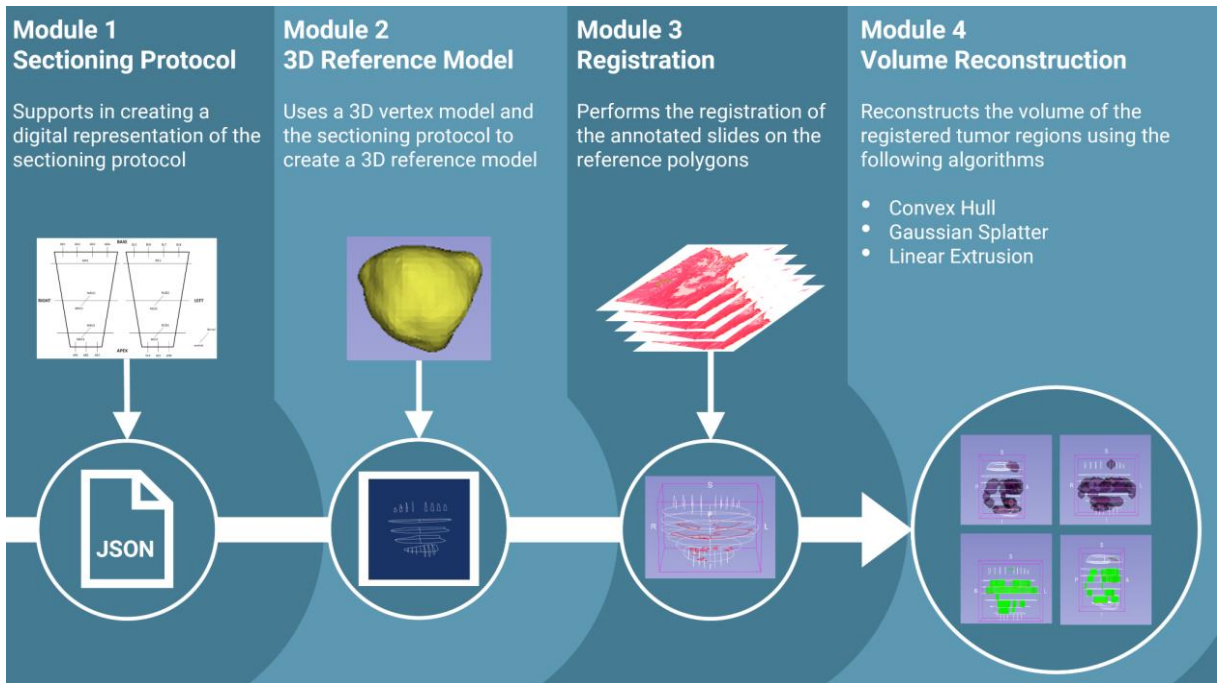

*Figure 1: Overview of the four different modules included in* 3D-SLIVER.

In the first module, the sectioning protocol of the prostatectomy specimen can be converted into the target JSON format using a graphical user interface (GUI). In the second module, this sectioning can be applied to an imported 3D model, yielding a set of polygons approximating the histologic slides. In the third module, the annotated tissue and tumor regions are fitted into these reference polygons. Finally, in the fourth module, the transformed tumor regions can be converted into volume reconstructions using three different algorithms. In the following, we describe the individual modules provided within 3D-SLIVER.

## Module 1: Sectioning Protocol

The sectioning protocol provides a schematic record of how the radical prostatectomy specimen was dissected, specifying the number of tissue blocks for each region as well as the corresponding tissue block identifiers. In most cases, the prostate is cut transversely into slices. These slices are often too large to fit on standard-sized glass slides, so they are usually divided into four pieces - but possibly into more, depending on the size of the slice. Apex and base are typically divided into right and left and then sagittally cut into several partial slices. Figure 2 shows an example of such a sectioning protocol.

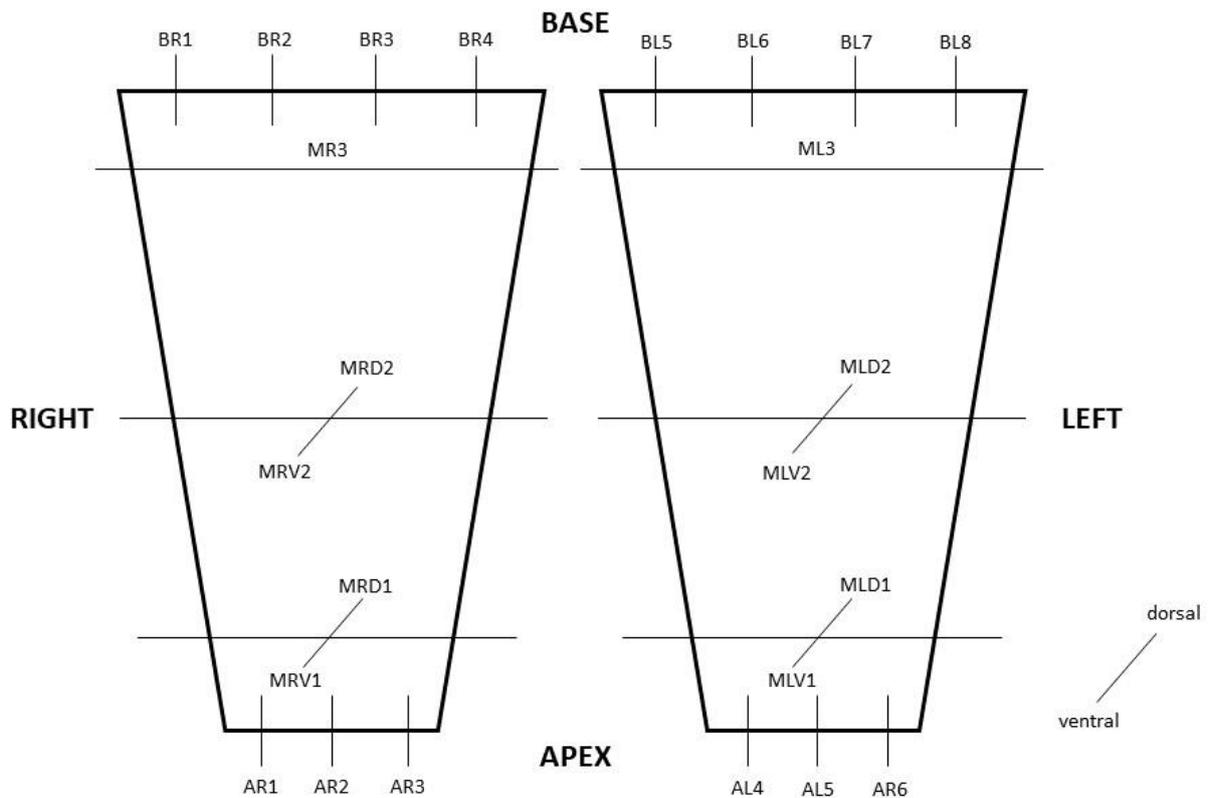

*Figure 2: Sectioning Protocol with the apex dissected into six sections and the base dissected into eight sections.*

Each letter represents one cassette of collected tissue, from which a paraffin block and, finally, the histological sections are created. For the apex and base, several pieces of tissue may be embedded within the same block. Thus, when creating the annotations (as described in Module 3: Registration), it is important to ensure that the individual tissue fragments are assigned to the correct polygons in the reference model.

The protocol can be digitized into a machine-readable JSON file via the Graphical User Interface (GUI) in the 3D Slicer module. In this JSON file, the sectioning protocol is divided into three different object types: apex, base and center slices. Apex and base are described using the same structure. They are each divided into a right and left segment, which are, in turn, divided sagittally into several fragments. Depending on the size of the prostatectomy specimen, it may be necessary to additionally divide the right and left sides along the frontal axis into ventral and dorsal fragments. This process is also covered by the JSON file structure. For each of these two (left and right) or four (left and right, each ventral and dorsal) sections of apex and base, the number of sagittally cut sections is specified, as well as a list of assigned IDs. The IDs consist of the block identifier, an L or R for the left or right side, an optional V or D for ventral or dorsal, and a sequential number, starting at 1.

For the central slices, the overall number of slices is specified, as well as a list of JSON objects for each individual slice. In each of these objects, it is specified whether the slices are subdivided into left and right only or additionally into dorsal and ventral, followed by a list of IDs containing the block identifier and the position.

## Module 2: 3D Reference Model

The reference model serves as the basis for registering (i.e., superimposing) the annotations made on digitized histological slides – so called whole slide images (WSIs), into the 3D space. It consists of polygons created by applying the sectioning protocol to a 3D surface model. Technically, this process is implemented using the Visualization Toolkit (VTK). To generate the reference polygons, the corresponding sectioning planes are determined from the JSON file describing the sectioning protocol, followed by computing their intersections with the 3D surface model. An example of the resulting reference polygons is given in Figure 3.

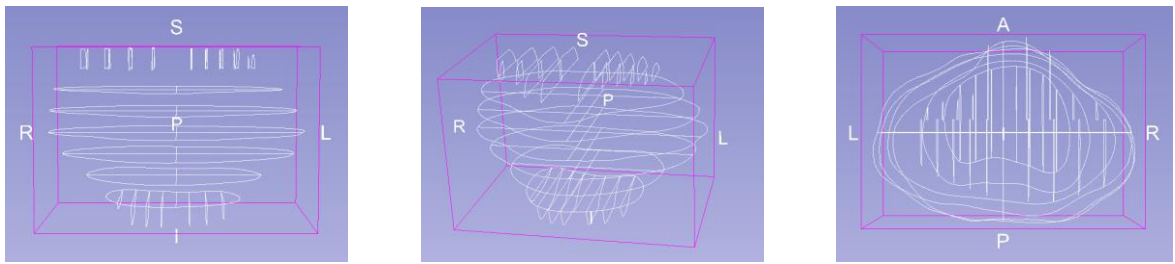

*Figure 3: Reference model of the prostate with approximated slide positions.*

Via the module's GUI, the file path to the sectioning protocol and the path to where the results should be saved can be specified. Furthermore, the GUI allows the specification of offset values for the apex and base to modify the thickness of these sections. To utilize this module, a 3D surface model must be imported into 3D Slicer. It can originate from different sources and must be available in OBJ, PLY, or STL format. The algorithm considers the size of the 3D model; therefore, the model may be arbitrarily scaled. However, depending on the source and data format, the model may be rotated differently than expected by the algorithm. Therefore, it may be necessary to rotate it before generating the reference polygons.

The results are stored in a single JSON file, consisting of a list of JSON objects, each with a name, a list of 3D point coordinates and a list of point pair indices forming the polygon's edges. The assignment to the WSI-based annotations is achieved via the name of the respective polygons as assigned during the creation of the sectioning protocol. Afterward, the results can be exported as OBJ files, each representing a single tissue slide of the reference model and stored in a specifiable directory. In contrast, the reference polygon ID is included in the filename. In the following, we describe several methods to generate a 3D surface model of the prostate as a basis for the reconstruction.

**3D Scanner**

A 3D scanner allows capturing an object's 3D shape and, depending on the device, its surface colors using different capturing techniques, such as lasers or structured light. For this purpose, the object is systematically sampled to determine the distance between the device and the points on the object's surface to reconstruct the 3D surface, yielding a point cloud, which can be transformed into a 3D surface model using Delaunay triangulation. During the development of 3D-SLIVER, we used the V2 3D scanner from Matter and Form[1], which is provided with a rotating table and height-adjustable lasers and measuring devices. We achieved the best results by shielding direct light using blackened cardboard, performing multiple scan iterations, and registering these results using the accompanying scanner software. When positioning the prostatectomy specimen at the standard position, the apex' surface could not be sampled sufficiently. This could be solved by elevating the tissue using a small, weighted metal rod plugged into the apical end of the urethra.

**MRI segmentation**

Another source to generate a 3D model of the prostate are MRI scans. In particular, the T2 sequence yields sufficient morphological details to thoroughly outline the organ's pseudo-capsule. 3D slicer already provides tools to segment anatomical structures in MRI scans and convert them into 3D models using Delaunay triangulation within the Segment Editor module. However, the deformations between MRI morphology and the fixed and sectioned tissue will be significantly more substantial compared to a scanned *ex vivo* specimen.

**Generic Model**

In some cases, it is not possible or desired to create a specific 3D model for every single case. This may be because no MRI images were taken, no 3D scanner may be available, or the reconstruction may be carried out retrospectively on previously sectioned prostate tissue. Therefore, a generic 3D model of an archetypical prostate is provided, as agreed on by two pathologists and two radiologists, created using the 3D modeling software Blender. The model was then distorted to maintain the same size along all dimensions (e.g., from left to right, from apex to base, and from dorsal to ventral). Although this may initially appear unrealistically, it allows to easily adjust the model to the specific dimensions of the prostate to be reconstructed.

## Module 3: Registration

Registration is a procedure in which two point sets (or images) are superimposed in the best determinable way using optimization algorithms. Polygons are point sets with edges between individual points, so they can be registered using the same algorithms. As the reference model consists of polygons approximating the histological slide contours, the registration can be performed between these polygons and the tissue contours. Subsequently, the transformation rule used to superimpose the section contour on the reference polygon can also be applied to the tumor regions in the section. This whole process requires a set of annotations to ensure the exact assignment of the tumor regions.

**Annotations**

---

[1] https://matterandform.net/scanner

To create the digital annotations of the tumor regions, the histological sections must first be scanned with a WSI scanner. Subsequently, the contours of the tissue and the respective tumor region of interest (ROI) must be annotated and stored as GeoJSON files, which can be achieved easily using QuPath [14].

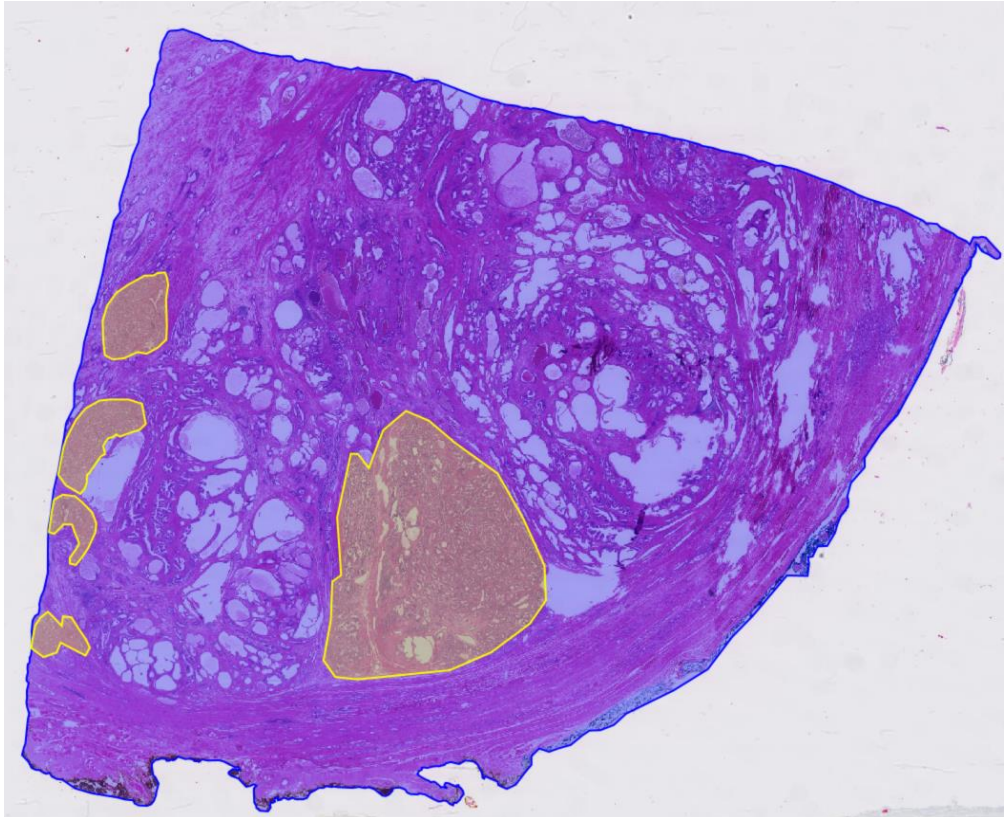

*Figure 4: Prostate tissue slide with ROIs for the contour in blue and five tumor ROIs in yellow.*

Two types of annotations need to be provided. First, an annotation of the tissue contour is required. Our tool expects the contour annotation in QuPath to be provided with a class named "Contour". In addition, further annotations can be provided to be transformed along with the contour, providing data for the 3D reconstruction. These can be, for example, tumor ROIs that are assigned the respective Gleason grade as a class. Any other structures can be annotated as well, as the annotation class allows for distinguishing between the structures in the 3D reconstruction. However, it is only necessary to provide contour annotations for those slides with structures to be reconstructed.

**Registration**

The main objective of the registration step is to superimpose the slide contours onto the 3D reference polygon in the best possible manner. Utilizing the polygons' points coordinates, the Coherent Point Drift Algorithm [15] is applied, preceded by a naive polygon upsampling algorithm to omit potential biases towards denser sampled parts of the polygon contours. Once the final transformation is established, it can then also be applied to the other annotated areas to correctly position them in the reference model and store the results in a JSON file.

For each annotation transformed in this way, the name of the respective WSI or, in the case of sections with multiple pieces of tissue, the name of the fragment, an ID for the case and, similar to the reference model, a list of coordinates and the edges connecting them are specified. In addition to the JSON file, the results can be bulk exported into one OBJ per slide or, alternatively, into one OBJ per transformed annotation per slide.

## Module 4: Volume Reconstruction

Finally, the registered tumor regions can be visualized in different ways. As such, they can be represented either as a convex hull around all existing polygons, as ellipticals based on Gaussian distributions using the Gaussian splatter algorithm, or with linear extrusion, an algorithm that orthogonally extrudes the contours of the registered regions. All of these methods serve to reconstruct the registered tumor regions in a 3D space. For an overview of the three-volume reconstruction methods, see Figure 5.

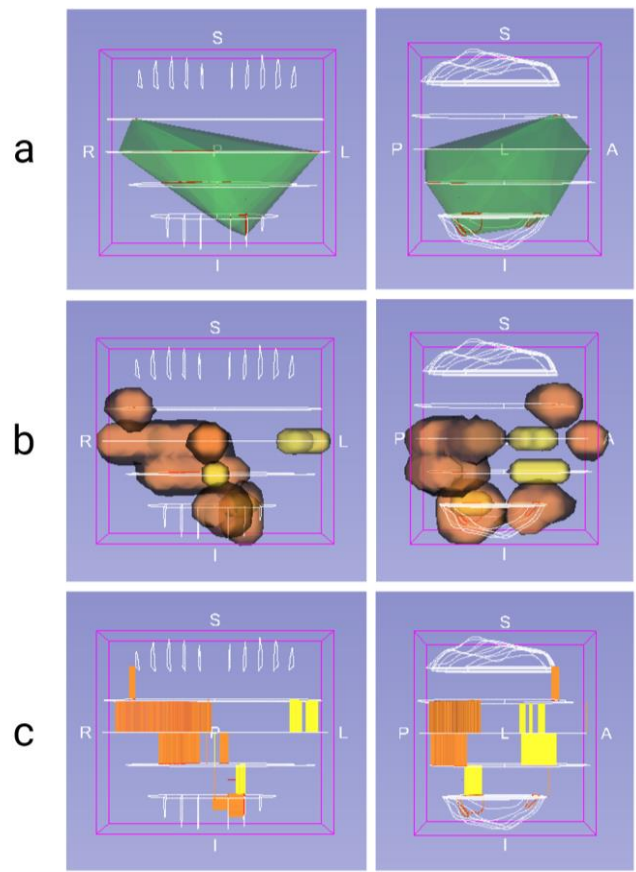

Figure X: Volume reconstruction algorithms of a prostatectomy specimen containing Gleason grades 3 and 4: a) Convex Hull (without class-specific colors) b) Gaussian Splatter c) Linear Extrusion

### Convex hull

The volume reconstruction with convex hulls creates a single shape enclosing all registered annotations and is intended to visualize the entirety of the tumor-affected area within the prostatectomy specimen. Specifically, it forms the smallest convex shape around all points of all registered annotations and, therefore, may also include tumor-free regions between different tumor foci or tumor protrusions. Hence, it is improper to approximate the tumor morphology, illustrate its focal distribution, or separate the reconstruction using different annotation classes (e.g., Gleason grades).

### Gaussian Splatter

The volume reconstruction with Gaussian Splatter creates a sphere-based, rather organic morphology based on the registered annotations. Each annotation is individually approximated with ellipticals by computing the Gaussian distribution within the annotation-specific point set. The respective mean and standard deviations along the three dimensions, as well as a configurable radius, determine the exact shape of the ellipticals. Contrary to the convex hulls, the Gaussian Splatter will consider the annotation classes and create one cohesive Gaussian Splatter for each of them. This approach is aimed at maintaining the overall tumor distribution without stressing annotation details.

### Linear extrusion

The volume reconstruction with Linear Extrusion takes the registered annotations and extrudes them to the approximated width of the transversal slices, as computed during the creation of the 3D reference model. As the sectioning for the apex and base differs from the sectioning of the mid-sections of the prostate, the extrusion is performed with regard to the specific location. Each registered annotation will produce precisely one Linear Extrusion. Similar to the Gaussian Splatter implementation, grouping by annotation class is performed if multiple classes are present. This approach aims to maintain the tumor distribution and annotation details as accurately as possible.

# Results

3D-SLIVER covers our complete process of reconstructing prostatectomy specimens in 3D. The four major steps are implemented as separate modules each providing a unique set of tools tailored towards their interplay as well to be flexibly replaced with our exported to additional software. This is mostly achieved by storing results as JSON or OBJ files.

For registration, the Coherent Point Drift algorithm was used in combination with the intersection over union (IoU) metric to determine the best one in a set of multiple registrations. In an exemplary examined case, an IoU of $0.91 \pm 0.04$ was achieved across all sections. To enhance the results, the number of polygon sampling points is increased before the actual registration. To determine a sufficient number, a test series was carried out on 31 polygon pairs from one prostatectomy specimen in which the number of sampling points was gradually increased from 100 to 1000, whereby stable IoU values could only be achieved across all sections from 400 points. By default, the registration module increases all polygons to at least 500 sampling points. To further improve the registration, a rough pre-alignment based on the polygons' center of gravity and scaling regarding the dimension with the greatest extent is introduced. To also account for the rotation, the polygon to be transformed was rotated at 8 different angles, each increasing bin steps of 45°, and finally, the registration result achieving the highest IoU value was used.

Three different volume reconstruction methods are implemented. Convex hulls enclose all registered annotations to visualize the entirety of the tumor-affected areas within the prostatectomy specimen. However, it loses information on morphology and density distribution and can not be used to illustrate different annotation classes. Gaussian splatter creates ellipticals based on the sampling points of each registered annotation to outline the overall tumor distribution and potentially smooth annotation details. Linear extrusion simply extrudes the annotations to the approximated width of the transversal slices to consider the annotation details as accurately as possible.

3D-SLIVER has been designed for flexible use and enables reconstruction even without a 3D scanner or MRI images. In the following, we show three use cases of how the tool can be used effectively.

**Use Case 1: Enhancement of Pathological Documentation**

This use case outlines how a 3D component can be incorporated into pathology reports. To keep the additional effort as little as possible, the generic prostate model, scaled to the measured dimensions of each specific prostatectomy specimen, is utilized. To digitize the sectioning protocol, the required information may be extracted from the pathology information system and automatically transformed into the correct JSON representation, if possible, or, alternatively, a laboratory technician may be entrusted with this task. Furthermore, the slide IDs of all slides containing tumor tissue should be stored. The polygons corresponding to these slides are then identified in the reference model, saved in a separate file and re-imported as tumor annotations. This way, neither annotations nor the registration procedure are necessary. Only the reference polygons are used to indicate the tumorous slides in the 3D reference model. This process requires little to no manual work, as ideally, every single step is automated, providing a simple and effective method to increase the information content of pathology reports.

**Use Case 2: Retrospective Evaluation of PI-RADS Predictions**

This use case outlines how the 3D reconstruction can be used to evaluate radiological assessments with the Prostate Imaging Reporting and Data System (PI-RADS). PI-RADS is a standardized framework facilitating the prediction of PCa in MRI images within a set of predefined regional zones. To evaluate these localized predictions, a correlation with histological findings is commonly used. This is achievable using established methods but requires several precautions in the pathological workflow, eliminating the possibility of retrospective application. However, this can be realized with 3D-SLIVER by generating the reference model from the MRI images, using linear extrusion to maintain the annotation details, and visualizing the results in combination with the PI-RADS predictions, allowing for a comprehensive assessment of the accuracy.

**Use Case 3: Statistical Modeling of the 3D Distribution of Different Morphological Patterns**

This use case outlines how the 3D distribution of PCa can be assessed using volume reconstruction. Several approaches have been described to investigate this, again involving varying amounts of manual effort. The necessary sub-step of transferring the pathological evaluation can be automated using 3D-SLIVER. Several approaches to this are possible. A standardized 3D model could be used to allow for easier computable distribution patterns. Additionally, a zonal distinction could be introduced by refining the generic 3D vertex model. Alternatively, the reconstruction could be based on MRI segmentations or 3D scanners to create a large database with the actual prostate shapes and the distribution within. This would result in an extensive database on the spatial distribution of PCa, laying the basis for further analysis.

# Discussion

This paper describes a comprehensive workflow for histological 3D volume reconstruction of prostatectomy specimens using different techniques. This workflow is implemented as an extension of the well-known software 3D Slicer, providing sub-modules for each major step in this process. The 3D-SLIVER extension is designed to be flexible in that various image acquisition sources can be used for the 3D models used to reconstruct the volume of the histological evaluation. Moreover, three distinct volume reconstruction algorithms are available to visualize different aspects.

There are many possible approaches to generate the 3D reference model of the prostate. With the 3D scanner, MRI and the generic vertex model, we have explored three possible sources, but the possibilities for alternatives are manifold. The model from the 3D scanner enables comparatively accurate capture of the fixed ex-vivo specimen [16], offering a solid basis for the registration of histological sections. However, if the histological evaluation is to be harmonized with the radiological findings, it is advisable to generate the reference model based on MRI images. Apart from the scaling with regard to the dimensions of the prostate, the generic model has only limited similarities with the actual tissue. A 3D scan of the surgical specimen is advantageous because the digitized tissue has already been partially processed so that the distortions caused by fixation have already occurred. In addition, the amount of information about the tissue's condition is significantly increased, which is particularly interesting to urologists. Instead of manually browsing the pathology report for indications of tissue damage or anomalies to draw conclusions about the surgical technique, the 3D model enables direct communication of abnormalities. Radiology could also benefit from this technology, for example, by deepening the understanding of preparation-related deformations [17]. In addition, the 3D model enables a more direct comparison with the MRI images, which can help with the interpretation and assessment of the images.

The generated 3D models differ significantly in their respective level of detail, depending on the imaging modality and methodology. However, since the processing for histology can result in significant tissue distortion, a higher level of detail in the 3D model does not necessarily improve the registration. Furthermore, the positions of the histological sections in the 3D reference model can only be approximated, so that more detail may even be counterproductive. For example, if there is a protuberance or indentation in the organ at the approximated position that is not apparent in the histological section, then this might even have a negative effect on registration. The selection of the method for generating the 3D reference model should therefore depend on the intended use. When feedback from radiology is desired, it appears reasonable to work with a segmented MRI. However, for feedback to urology, a choice between 3D scanner, photogrammetry or generic model can be made, depending on the available technology, time and expertise.

Coherent Point Drift was chosen as the registration algorithm so that arbitrary 3D surface models can be used as a reference. Therefore, while our approach is flexible with respect to data sources, it suffers in accuracy, since tissue structures cannot be taken into account in the registration process. An open question is whether the registration could be further improved if whole-mount slides were obtained from the tissue and digitized with appropriate scanners, since this would eliminate further subdivision and, thus, distortion of the tissue.

Three distinct methods for volume reconstruction based on annotations are provided within 3D-SLIVER: Convex hull, Gaussian Splatter, and Linear extrusion. With the convex hull, a single shape enclosing all annotations, providing an overview of the entire tumor-affected area is achieved, which, in turn, is inappropriate for illustrating the tumor's spatial distribution. Gaussian Splatter creates rather organic shapes based on annotations, considering annotation classes for cohesive representation without emphasizing annotation details. Linear extrusion extrudes annotations to the approximated width of transversal prostate slices, maintaining tumor distribution and annotation details with the highest precision. Each method offers unique advantages and limitations, contributing to the understanding of PCa visualization techniques.

The tool enables a significant improvement in the flow of information, which was previously limited to the pathology report and, if applicable, the MRI images and radiological findings. As shown in Figure 6, the tool enriches the flow of information in that information from the diagnostic process in pathology can be shared transparently with the other specialist disciplines involved.

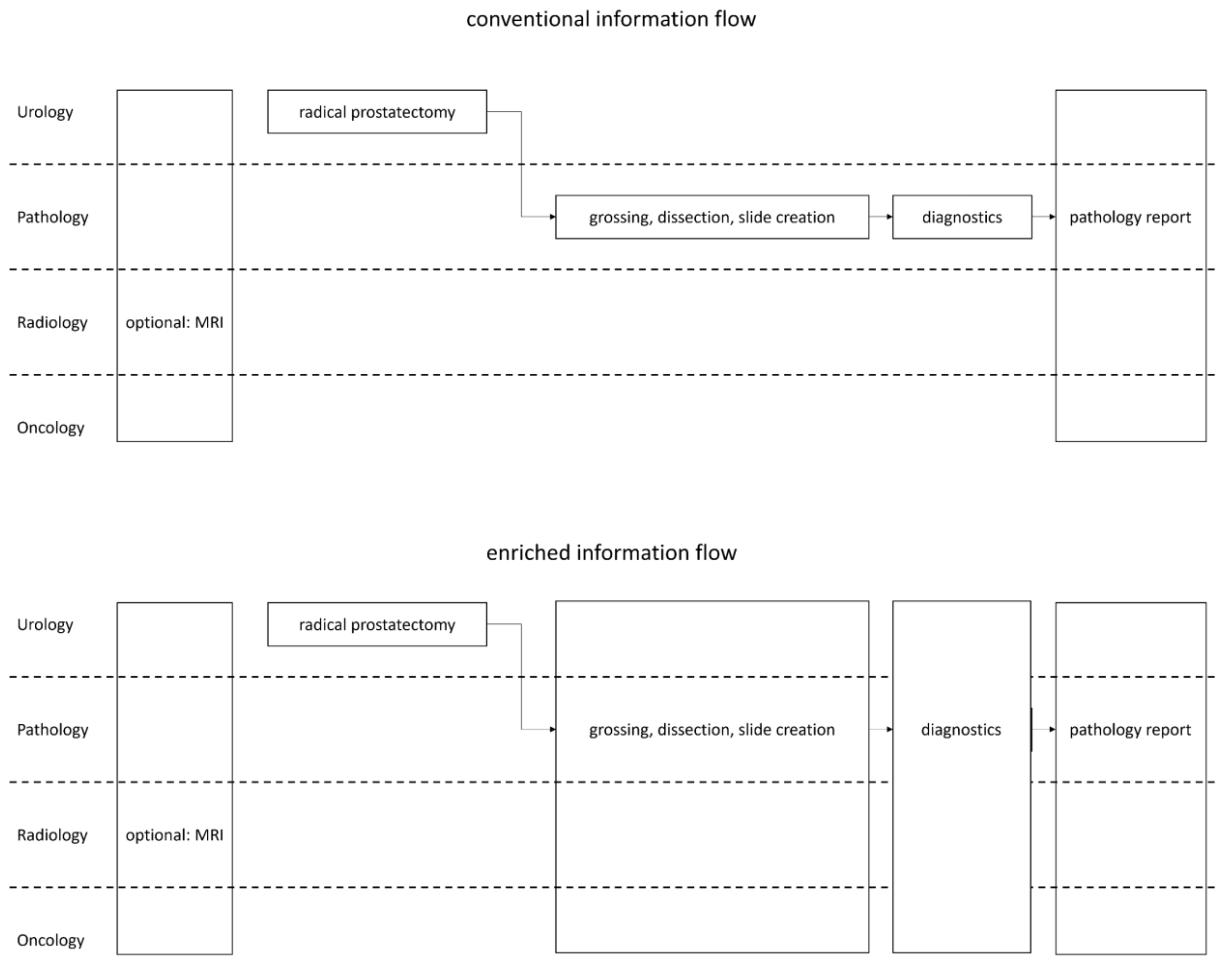

*Figure 6: Conventional information flow compared to the flow enriched with information from the 3D reconstruction.*

3D volume reconstruction of the tumor foci is highly relevant for radiology and oncology. Radiologists can use it to directly compare their interpretations of the MRI data by projecting the histopathological evaluation into the MRI data [18]. If PI-RADS is used, a comparison can be made for each individually assessed lesion or PI-RADS zone. Oncology could benefit in the long term by adding a more refined spatial component to the prognosis.

The proposed workflow offers notable advantages, requiring minimal manual intervention and the capability to be executed independently from the diagnostic workflow. It facilitates a deeper comprehension of tumor growth patterns, potentially enhancing treatment and diagnostic strategies. However, its accuracy falls short compared to alternative approaches. When surgical procedures, including frozen sectioning during the operation, such as NeuroSAFE, are performed, 3D reference models from MRI or the provided generic model would not adequately represent the prostatectomy specimen. In this case, the described workflow is limited to 3D models directly derived from the ex vivo specimen (e.g., using a 3D scanner). Generally, the workflow presented here cannot be expected to provide high-precision reconstruction at a microscopic level. However, it can be helpful in interdisciplinary communication, e.g., when giving feedback to radiology or urology. Pathological documentation could be enriched, introducing new dimensions for quality assurance and refining the means of obtaining Gleason grade percentages and estimating size and localization, lining up with some of the objectives pursued with synoptic reporting [19]. Emerging technologies that enable high-throughput 3D image acquisition, such as light sheet microscopy, optical coherence tomography, or micro-CT, hold promise for enhancing the histological 3D reconstruction of prostatectomy specimens. They have the potential to serve as a bridge between macroscopic and microscopic morphology, improving the accuracy and fidelity of diagnostic imaging in diagnostic routine.

# Conclusion

Here, we present a method for 3D reconstruction of radical prostatectomy specimens for interdisciplinary communication with radiology and urology. In contrast to existing solutions, this approach is low-threshold in application, can be used without additional equipment apart from a whole slide scanner, performed retrospectively and used in the context of documentation and quality assurance.

# Declaration of conflicting interests

The authors declare that no potential conflicts of interest exist with respect to the research, authorship, and/or publication of this article.

# Funding

This work was funded by the German Federal Ministry for Economic Affairs and Climate Action (BMWK) [Project "EMPAIA"; FKZ 01MK20002] and by the German Federal Ministry of Education and Research (BMBF) [Project "PROSurvival"; FKZ 01KD2213].

# Footnote

Guarantor TB

# Appendix

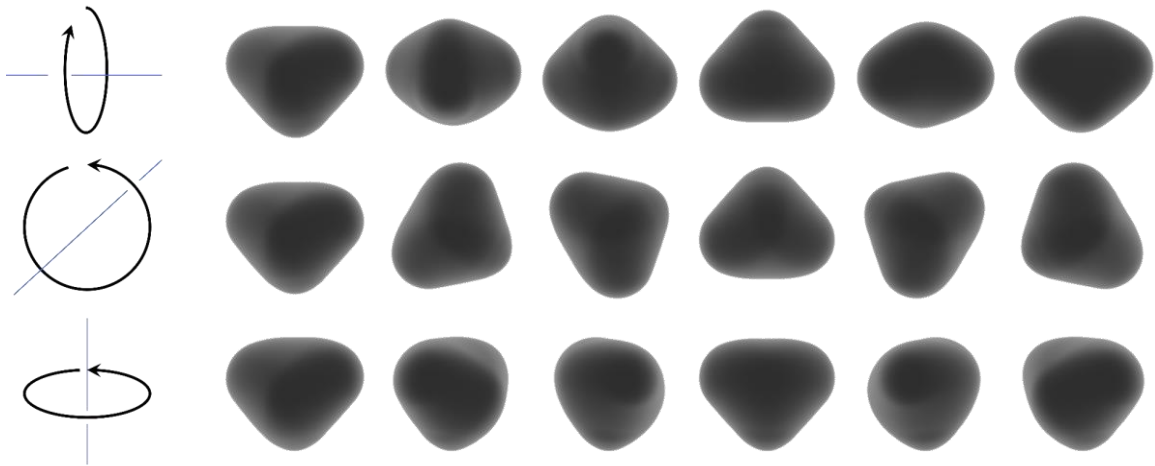

Appendix 1: Rendering of the generic model of the prostate with an equal size along all dimensions.